\documentclass[11pt]{article}

\usepackage[final]{acl}

\usepackage{times}
\usepackage{latexsym}
\usepackage{amsmath}

\usepackage{algorithm}
\usepackage{algorithmic}

\usepackage{tcolorbox}
\usepackage{parskip} 

\usepackage[T1]{fontenc}
\usepackage{tabularx}
\usepackage{booktabs}
\usepackage{array}
\usepackage{multirow}
\usepackage{caption}
\usepackage{pdflscape}
\usepackage{pgfplots}
\usepgfplotslibrary{groupplots}
\usepackage{afterpage}
\usepackage{arydshln}
\usepackage{graphicx}
\usepackage{datatool}
\usepackage{tikz}
\usepackage{amssymb}
\usepackage{enumitem}
\pgfplotsset{compat=1.17}
\usepackage{subcaption}
\usepackage{listings}
\usepackage{xcolor}
\lstdefinestyle{mypython}{
  language=Python,
  basicstyle=\ttfamily\small,
  numbers=left,
  numberstyle=\tiny\color{gray},
  numbersep=8pt,
  breaklines=true,
  breakindent=0pt,
  keywordstyle=\color{blue},
  stringstyle=\color{orange},
  commentstyle=\color{gray},
  showstringspaces=false,
  columns=flexible,
  tabsize=2,
  xleftmargin=1.5em,
  framexleftmargin=1.5em,
  backgroundcolor=\color{white}
}
\usetikzlibrary{patterns}
\usepackage[title]{appendix}

 \title{MountainLion: A Multi-Modal LLM-Based Agent System for Interpretable and Adaptive Financial Trading}

\author{
  \textbf{Siyi Wu}$^{1}$, \textbf{Junqiao Wang}$^{2}$, \textbf{Zhaoyang Guan}$^{3}$, \textbf{Leyi Zhao}$^{4}$, \textbf{Xinyuan Song}$^{5}$, \textbf{Xinyu Ying}$^{6}$, \textbf{Dexu Yu}$^{7}$ \\
  \textbf{Jinhao Wang}$^{7}$, \textbf{Hanlin Zhang}$^{8}$, \textbf{Michele Pak}$^{9}$, \textbf{Yangfan He}$^{10}$, \textbf{Yi Xin}$^{11}$, \textbf{Jianhui Wang}$^{12}$, \textbf{Tianyu Shi}$^{13}$\thanks{Corresponding author. \url{tys@cs.toronto.edu}} \\
  $^{1}$ The University of Texas at Arlington  \quad
  $^{2}$ Sichuan University \quad
  $^{3}$ Northwestern University \quad \\
  $^{4}$ Indiana University \quad 
  $^{5}$ Emory University \quad 
  $^{6}$ Nankai University \quad 
  $^{7}$ MountainLion Research \quad \\
  $^{8}$ Xi'an University of Electronic Science and Technology \quad 
  $^{9}$ Kyoto University \quad \\
  $^{10}$ University of North Carolina at Chapel Hill \quad 
  $^{11}$ Nanjing university \quad \\
  $^{12}$ Tsinghua University \quad
  $^{13}$ University of Toronto 
}

\begin{document}
\maketitle

\begin{abstract}
Cryptocurrency trading is a challenging task requiring the integration of heterogeneous data from multiple modalities. Traditional deep learning and reinforcement learning approaches typically demand large training datasets and encode diverse inputs into numerical representations, often at the cost of interpretability. Recent progress in large language model (LLM)-based agents has demonstrated the capacity to process multi-modal data and support complex investment decision-making. Building on these advances, we present \textbf{MountainLion}, a multi-modal, multi-agent system for financial trading that coordinates specialized LLM-based agents to interpret financial data and generate investment strategies. MountainLion processes textual news, candlestick charts, and trading signal charts to produce high-quality financial reports, while also enabling modification of reports and investment recommendations through data-driven user interaction and question answering. A central reflection module analyzes historical trading signals and outcomes to continuously refine decision processes, and the system is capable of real-time report analysis, summarization, and dynamic adjustment of investment strategies. Empirical results confirm that MountainLion systematically enriches technical price triggers with contextual macroeconomic and capital flow signals, providing a more interpretable, robust, and actionable investment framework that improves returns and strengthens investor confidence.
\end{abstract}

\section{Introduction}

Cryptocurrency trading requires fast, explainable decision-making across multiple modalities—including real-time market data, news narratives, social signals, and historical indicators~\cite{liu2022survey}. Traditional deep learning (DL) and reinforcement learning (RL) approaches encode such data into latent numeric vectors, often sacrificing transparency and adaptability~\cite{chen2020predicting, nakamoto2008bitcoin}. Moreover, most systems rely on static architectures that lack responsiveness to evolving market sentiment and regulatory context.

Recent progress in Large Language Models (LLMs), particularly when paired with Retrieval-Augmented Generation (RAG)\cite{lewis2020retrieval}, has opened new directions for grounded, real-time, and interpretable financial reasoning. LLM-based financial systems such as\cite{zhang2021btctrader} have explored combining technical indicators with LLM-driven summarization. Moreover, the introduction of LLM agents~\cite{xu2023survey} has further strengthened the capabilities of financial systems by enabling dynamic, interactive, and multi-turn analysis pipelines. In particular, financial report generation can benefit from this agent-based framework, which supports continuous refinement through iterative, dialogue-like reasoning~\cite{wu2023survey}. This agent-based paradigm not only facilitates the integration of diverse data sources but also enhances the accuracy and diversity of investment strategies~\cite{wu2023financial}.

\begin{figure}[!ht]
    \centering
    \includegraphics[width=\linewidth]{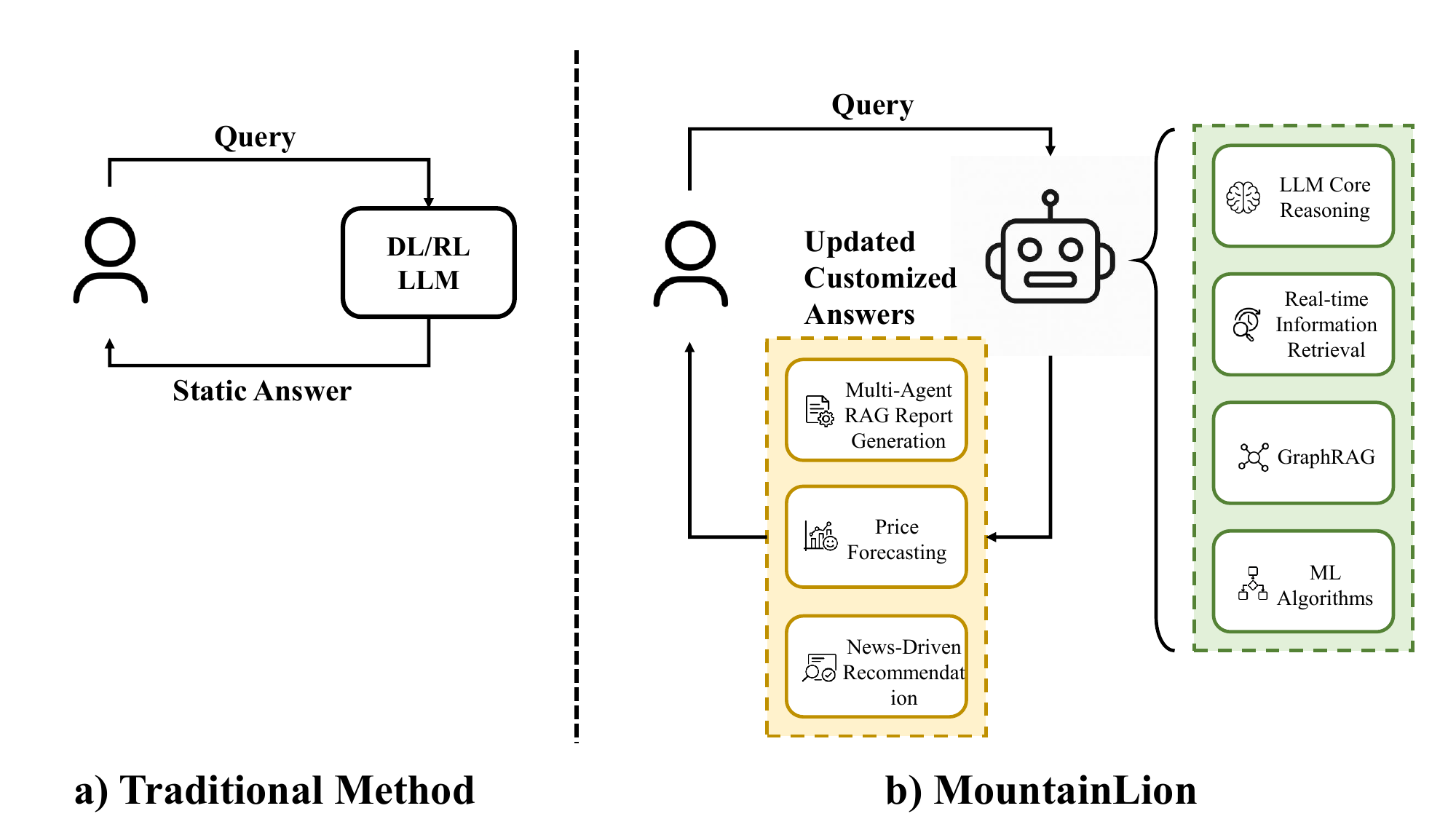}
    \caption{Comparison of (a) traditional DL/RL-based static QA pipelines and (b) our MountainLion framework. MountainLion produces updated, personalized answers through multi-agent collaboration, RAG, and reflective reasoning.}
    \label{fig:intro_comparison}
\end{figure}

In this work, we introduce \textbf{MountainLion}, a multi-agent, RAG-enabled financial analysis framework tailored to cryptocurrency markets. Traditional QA-style pipelines typically employ static deep learning or reinforcement learning models that return fixed, predetermined answers to each query~\cite{mnih2015human,seo2016bidirectional}. In contrast, MountainLion dynamically generates customized and up-to-date responses by leveraging multi-agent collaboration, real-time information retrieval, and graph-based reasoning. Its reflective, modular architecture actively synthesizes personalized answers, supported by a RAG backbone and coordinated agent-based interactions. Figure~\ref{fig:intro_comparison} highlights this contrast.

MountainLion consists of four core modules:

\begin{itemize}[left = 0em]
    \item \textbf{User Interface (UI)} provides an interactive dashboard where users can receive alerts, customize preferences, and monitor system outputs.
    \item \textbf{Core Business Logic} handles multi-agent report generation, price forecasting, and news-driven recommendation services.
    \item \textbf{AI Engine} includes LLM-based reasoning agents, real-time retrieval modules, graph-based RAG (GraphRAG)~\cite{sun2023graphrag} components, and ML-based predictive models.
    \item \textbf{Database Layer} integrates data from exchanges, news aggregators, and on-chain metrics, storing structured outputs like predictions and reports.
\end{itemize}

Within this architecture, our system deploys specialized LLM agents assigned to roles such as news analysis, chart interpretation, sentiment assessment, and KOL (Key Opinion Leader) tracking. These agents collaborate through a shared reflection module that evaluates the accuracy and consequences of prior outputs to refine future reasoning. RAG modules provide access to external, up-to-date information during inference, reducing hallucination and improving response quality.

Extensive experiments and ablation studies demonstrate that visual reflection not only boosts medium-term forecasting accuracy but also supports high-quality financial report generation with iterative refinement through multi-turn reasoning. By accommodating flexible investment horizons, the system enables short-term execution triggers, mid-term allocation planning, and long-term portfolio management within a unified, agent-based, interpretable framework that can scale to real-world cryptocurrency trading needs.

\section{Related Work}

\paragraph{LLMs in Financial Forecasting.}
Recent efforts have combined LLMs with financial analytics for enhanced interpretability. \cite{zhang2021btctrader} explore LLM summarization for price classification, overbought/oversold detection, and technical crossover detection. \cite{li2024cryptotrade} propose a reflection-based LLM trading agent. \cite{wu2023financial,yang2023agent} further highlight LLM agents for financial dialogue and benchmarking. However, these systems primarily rely on textual inputs, lacking integration with real-time market dynamics and visual data.

\paragraph{Sentiment-Aware Modeling.}
Extracting and interpreting investor sentiment from social media platforms like Twitter/X and Reddit has proven valuable. \cite{alonso2020cryptoBERT} introduced CryptoBERT, while \cite{xu2021perplexitynews} employ perplexity scoring for relevance ranking. \cite{si2021survey,ma2023smfinbert} further explore advanced sentiment models in social financial contexts. Our system expands on this by incorporating influencer (KOL) tracking and sentiment consistency verification using bot detection and engagement analysis.

\paragraph{Adaptive News and Regulation Awareness.}
\cite{liu2021cryptoalert} explore named entity recognition and keyword filtering for financial event extraction. While effective, most lack contextual reasoning or real-time annotation. \cite{yang2021finbert,wang2023multilingual} address multilingual and prompt-based financial event understanding. We incorporate multilingual support and RAG-enabled semantic filtering, enhancing interpretability of evolving regulatory narratives~\cite{lee2022strategic}.

\paragraph{Investment Strategy and Portfolio Optimization.}
Medium-to-long-term strategy modeling has focused on macro signals and risk-balanced allocation~\cite{li2023macromodels}. Dollar-cost averaging (DCA) and event-triggered rebalancing show promise~\cite{lee2022strategic}. \cite{yang2020deepportfolio,shen2021multiagent} further explore portfolio optimization through deep learning and multi-agent methods. MountainLion adopts a dual-agent collaboration framework to align reports with investment horizons and hedge exposure via inverse correlation modeling.

\paragraph{On-Chain and Visual Signals.}
Studies like \cite{nguyen2023volatility} and industry analysis highlight the predictive strength of wallet movements and miner flows. \cite{chen2020bitcoin,liu2023blockchain} further discuss on-chain activity analytics for volatility and risk forecasting. Our system utilizes OHLC data~\cite{sezer2020financial} across multiple granularities (1d, 1h, 5m), stored in a structured MySQL database for longitudinal forecasting~\cite{kim2019cnnforecast}.

\section{System Design}

\textbf{MountainLion} is a modular and extensible framework for quantitative investment analysis in the Web3 ecosystem. It addresses key challenges in cryptocurrency analytics by integrating diverse data sources, real-time market data, news sentiment, and on-chain activity, into a unified AI-driven decision-making pipeline. The platform provides data-driven insights to support trend forecasting, risk evaluation, and strategic allocation.

\subsection{Architecture Overview}

\begin{figure}[!ht]
\centering
\includegraphics[width=1\linewidth]{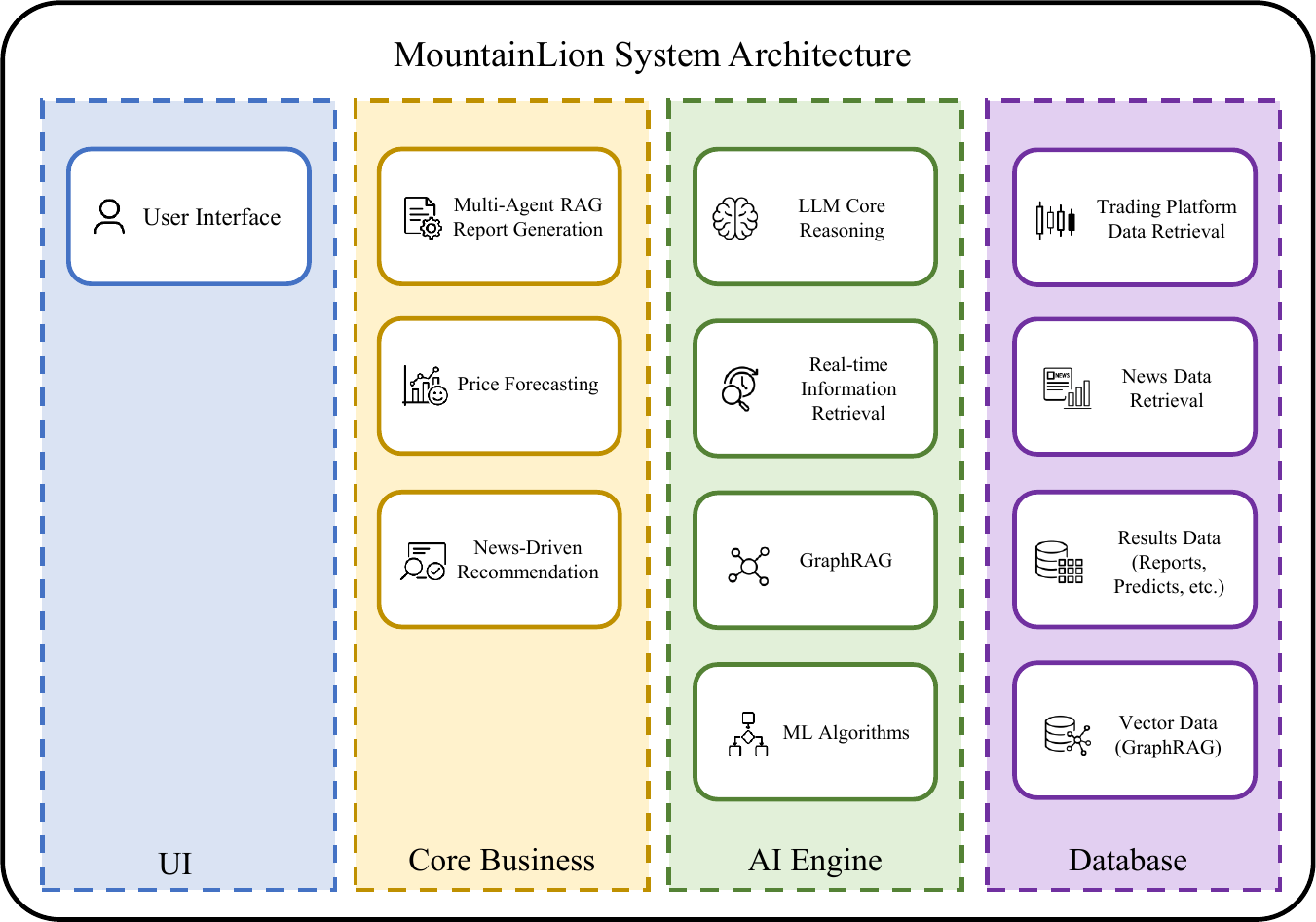}
\caption{MountainLion system architecture: modular design across UI, Core Business, AI Engine, and Database layers.}
\label{fig:architecture}
\end{figure}

MountainLion adopts a layered architecture designed for modularity. Its four principal layers: User Interface, Core Business, AI Engine, and Database, are depicted in Figure~\ref{fig:architecture}.

\textbf{User Interface Layer}: The UI provides a multilingual, real-time interaction dashboard supporting both chat and visual analytics.

\textbf{Core Business Layer} It implements domain-specific functionalities that orchestrate financial reasoning. It combines a multi-agent RAG-based report generation pipeline~\cite{yang2023agent} with modules for price forecasting and news-driven recommendations~\cite{lewis2020retrieval}. By blending technical signals, sentiment analysis, and fundamental macroeconomic perspectives, this layer ensures that generated investment strategies adapt to evolving narratives and user contexts across multiple time horizons.

\textbf{AI Engine Layer} It serves as the reasoning backbone, featuring an LLM-powered coordination module, continuous real-time retrieval from exchanges and news sources, and a GraphRAG mechanism~\cite{peng2024graph} for graph-structured semantic enrichment. Complemented by classical statistical and machine learning models, this layer supports robust, explainable, and data-driven predictions, addressing both interpretability and adaptability.

\textbf{Database Layer} It provides scalable storage and retrieval for trading data, news articles, generated reports, and dense semantic embeddings. This ensures that MountainLion maintains consistency, supports traceable workflows, and can efficiently handle the diverse data types required for modern cryptocurrency investment analytics.

\subsection{Report Generation and Optimization}

\begin{figure}[!ht]
\centering
\includegraphics[width=1\columnwidth]{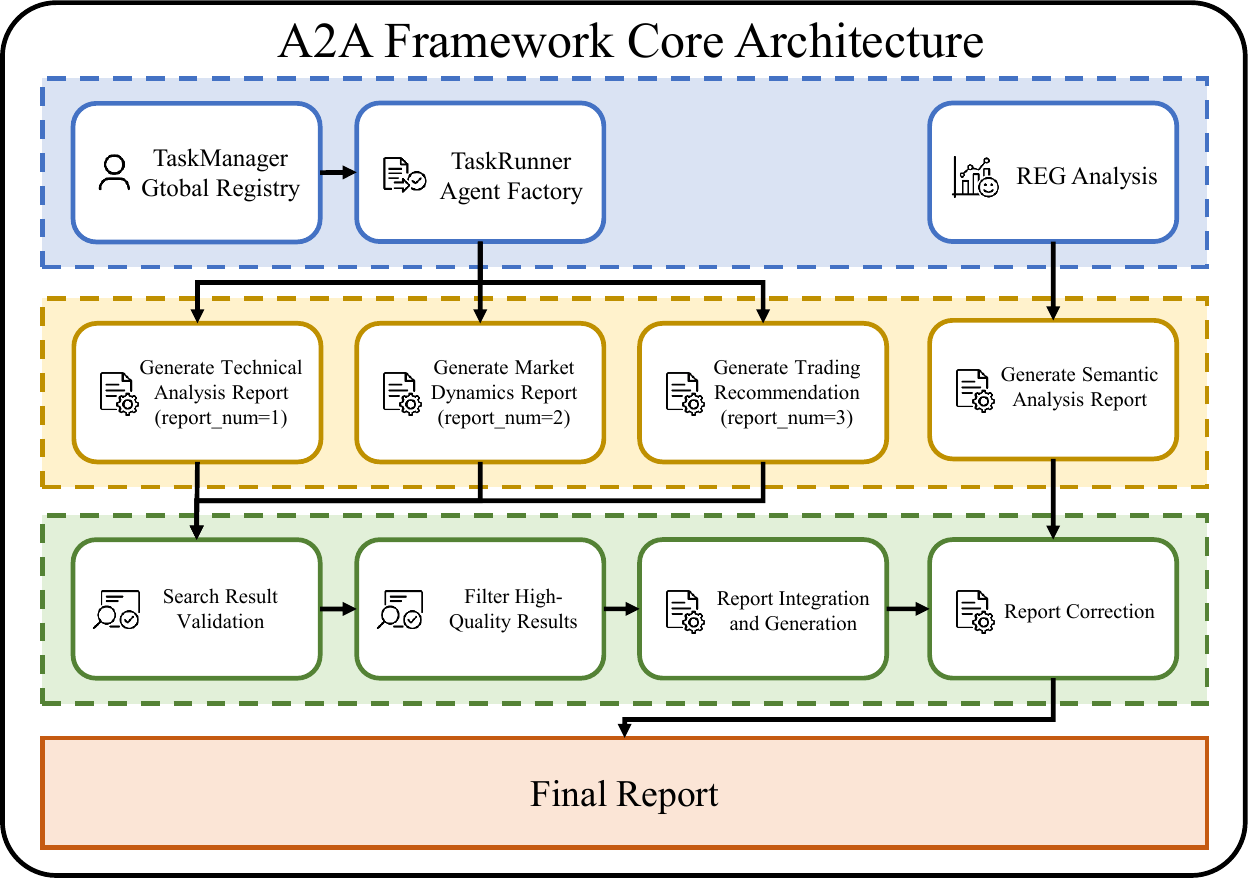}
\caption{Overview of the Report Generation Pipeline}
\label{fig:report_gen}
\end{figure}

To generate context-aware digital asset reports, \textbf{MountainLion} implements a structured multi-stage pipeline that coordinates agent collaboration, signal validation, and semantic refinement. As shown in Figure~\ref{fig:report_gen}, the process consists of: (1) task decomposition and specialization, (2) parallel analysis, and (3) report synthesis and enhancement.

\subsubsection{Agent-Based Task Decomposition}

A user-defined report request \( \mathcal{T} \) is first classified and routed to specialized expert agents \( \{A_1, A_2, A_3, A_4\} \). The \textbf{Technical Analysis Agent (\(A_1\))} Processes historical price and volume data to compute classical technical indicators (e.g., RSI, MACD, Bollinger Bands). Outputs include support/resistance zones and volume trends. Next, the \textbf{Market Dynamics Agent (\(A_2\))} Synthesizes external signals such as real-time news, capital flow, and sentiment indexes. Then, the \textbf{Trading Recommendation Agent (\(A_3\))} Integrates outputs from \( A_1 \) and \( A_2 \), generating multi-horizon trading strategies across different timescales. Finally, the \textbf{Semantic Agent (\(A_4\))} Refines the combined outputs through LLM-based analysis, enhancing coherence, logical flow, and lexical consistency.

\subsubsection{Report Enhancement and Optimization}
To ensure reliability, each agent $A_i$ formulates an information retrieval query \( Q_i \), which is evaluated based on relevance, recency, and source credibility. The validated signals resulting from each query, denoted as \( \mathcal{S}_{Q_i}^\star \), are then synthesized into the agent-specific partial report \( \mathcal{R}_i \). These partial reports are integrated through a centralized function to produce a consolidated draft report:
\begin{equation}
\mathcal{R}_{\text{raw}} = f_{\text{integrate}}(\mathcal{R}_1, \mathcal{R}_2, \mathcal{R}_3, \mathcal{R}_4).
\end{equation}
To align this draft report with prevailing market conditions, a final-stage enhancement is applied using a Perplexity-based retriever. Specifically, external signals \( \mathcal{S}_{\text{pplx}} \) are retrieved through prompts derived from the content of \( \mathcal{R}_{\text{raw}} \), yielding the final enhanced report:
\begin{equation}
\mathcal{R}_{\text{enhanced}} = f_{\text{augment}}(\mathcal{R}_{\text{raw}}, \mathcal{S}_{\text{pplx}}).
\end{equation}
Intermediate partial reports \( \mathcal{R}_i \) are cached following time-sensitive policies, with freshness checks conducted prior to re-execution to avoid redundant computation and maintain responsiveness.

The full formulation of agent-specific reasoning functions, along with their intermediate representations, scoring criteria, prompt construction methodology, and caching strategy, is detailed in Appendix~\ref{appendix:report-details}. This multi-agent, retrieval-enhanced framework ensures that MountainLion generates investment reports that are timely, data-grounded, and semantically robust.

\subsection{Price Forecasting}

\begin{figure}[!ht]
  \centering
  \includegraphics[width=1\columnwidth]{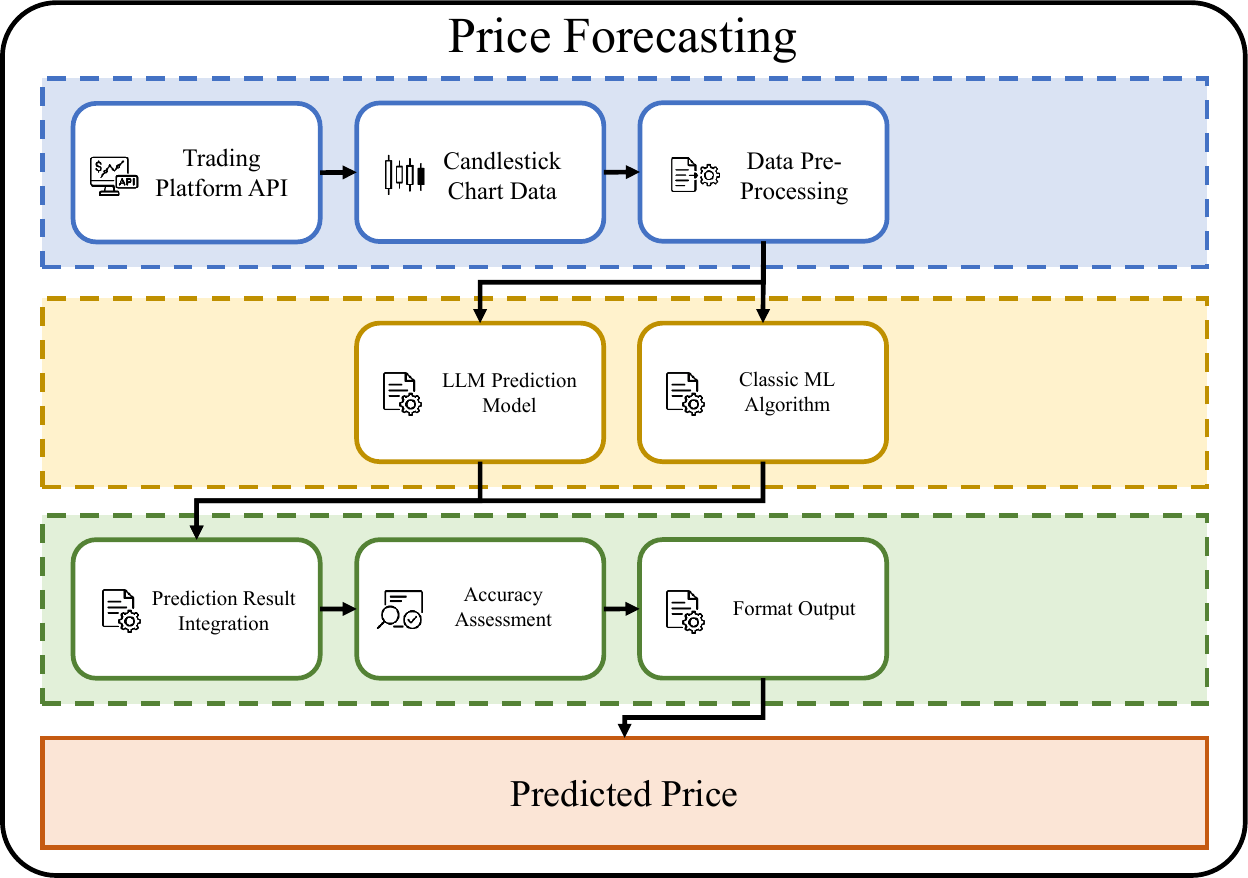}
  \caption{Architecture of the Price Forecasting System.}
  \label{fig:forecast-pipeline}
\end{figure}

To support timely investment decisions and enhance the credibility of financial analyses from a statistical perspective, \textbf{MountainLion} adopts a dual-path price forecasting framework that combines statistical learning with LLM-based reasoning. The forecasting module operates two coordinated parallel tracks to generate timely and reliable price predictions across multiple granular time horizons, denoted as $\mathcal{T} \in \{1\text{d}, 1\text{h}, 5\text{min}\}$, as illustrated in Figure~\ref{fig:forecast-pipeline}.

First, the LLM-Based Forecasting path processes structured OHLCV (open, high, low, close, volume) data alongside sentiment embeddings extracted from financial news, allowing the language model to generate multi-step forecasts conditioned on both quantitative and qualitative signals.  

Second, the ML-Based Forecasting path applies classical machine learning models, such as ridge regression and decision trees, to engineered technical features, providing lightweight and low-latency predictive outputs.  

The outputs from these two forecasting tracks, denoted as $\hat{Y}_{\text{LLM}}$ and $\hat{Y}_{\text{ML}}$, are aligned and integrated through a weighted fusion function:
\begin{equation}
\hat{Y}_{\text{final}} = f_{\text{fusion}}(\hat{Y}_{\text{LLM}}, \hat{Y}_{\text{ML}})
\end{equation}
where the weights are adaptively updated according to the rolling historical accuracy of each predictor.  

Forecast evaluation is based on two principal criteria: absolute accuracy, which measures the magnitude of prediction error, and directional correctness (win rate), which quantifies trend alignment. To enhance interpretability for end users, a template-conditioned text generator $g_{\text{text}}$ automatically converts these numerical forecasts into natural language summaries tailored for investor communication.  

Each forecast horizon $\mathcal{T}$ defines the prediction granularity, for example using two future candles for daily forecasts, or 24 steps for five-minute intervals, as discussed by Bandara et al.~\cite{bandara2020lstm}. All prediction results are systematically archived in resolution-specific tables to facilitate evaluation and downstream decision modules.  

Full specifications of input data formats, mathematical derivations of the fusion and weighting functions, and details of the adaptive performance adjustment loop are provided in Appendix~\ref{appendix:forecast-details}.

\subsection{News-Driven Recommendation System}

\begin{figure}[!ht]
  \centering
  \includegraphics[width=\columnwidth]{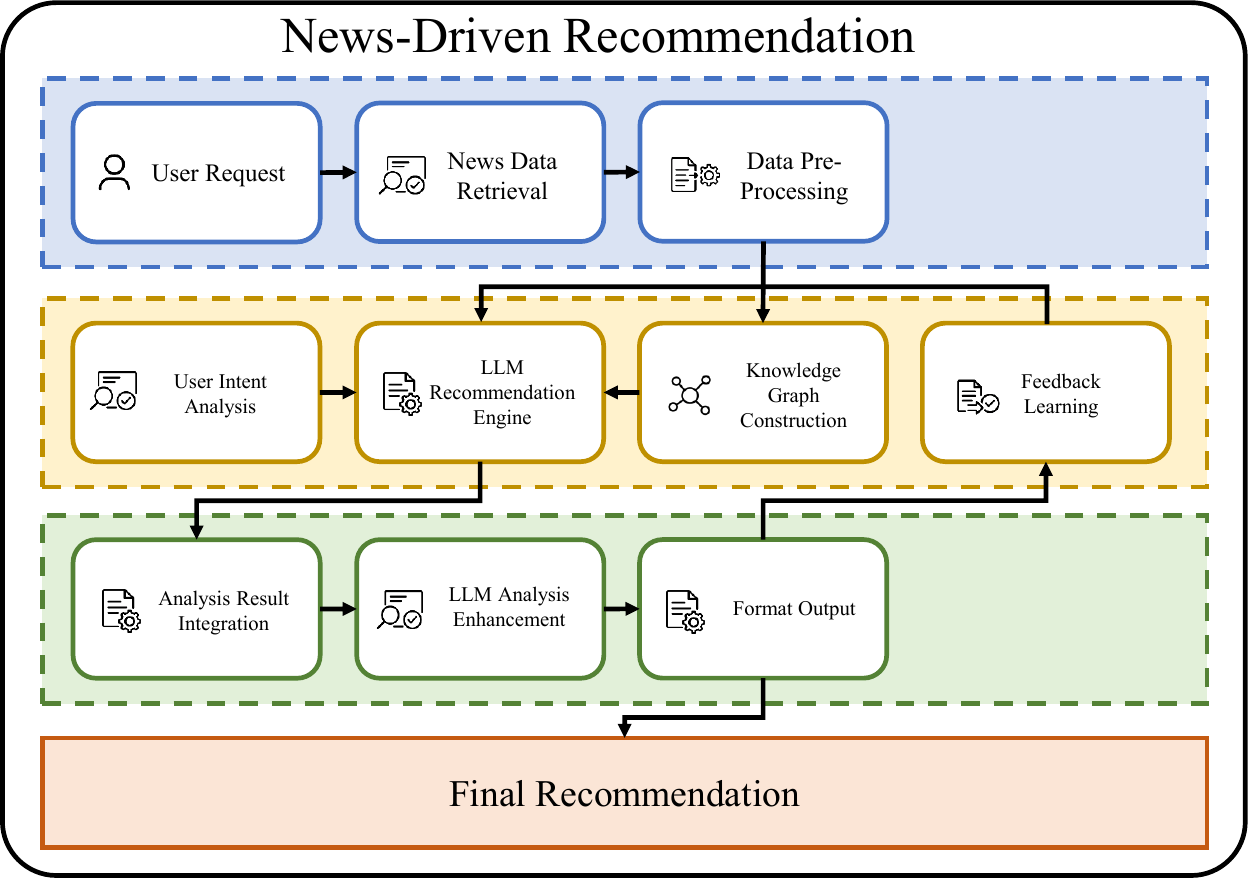}
  \caption{Architecture of the News-Driven Recommendation System}
  \label{fig:news-reco-pipeline}
\end{figure}

To support sentiment-aware investment decisions, \textbf{MountainLion} incorporates a news-driven recommendation engine that fuses user intent, news semantics, and graph-based reasoning. As shown in Figure~\ref{fig:news-reco-pipeline}, the system integrates multi-source information retrieval, contextual entity analysis, and LLM-based summarization to produce interpretable trading insights.

\subsubsection{News Processing and Semantic Enrichment}

\textbf{MountainLion} continuously ingests financial news from APIs, RSS feeds~\cite{orr2003rss}, and curated portals to enrich its market situational awareness. A semantic annotation module classifies each article according to sentiment polarity (for example, \texttt{Bullish} or \texttt{Bearish}) and extracts salient entities, such as tokens, events, and organizations, via named entity recognition (NER) techniques~\cite{li2022survey}. These annotations allow the system to build structured news representations that support graph-based modeling, clustering, and time-sensitive filtering aligned with the user’s investment horizon.  

A dynamic knowledge graph~\cite{hogan2021knowledge} is constructed to encode semantic relationships among news items and extracted entities. In this graph, nodes represent either documents or named entities, while edges capture co-occurrence or context-derived associations. This graph structure supports multi-hop reasoning, enabling grounded summarization of news content and evidence-backed investment recommendations.  

When a user submits a query, the system infers implicit preferences, including target asset categories, risk appetite, and investment time frames, and uses these to compose retrieval prompts. An instruction-tuned LLM then generates investment recommendations by integrating evidence from retrieved news documents and the graph context.  

Finally, user engagement signals, such as click-through rates and feedback ratings, are continuously logged and incorporated into a lightweight feedback policy module, like in~\cite{joachims2002optimizing}, which refines both retrieval and summarization strategies over time. The formal definitions, graph construction methodology, query parsing framework, and associated optimization procedures are detailed in Appendix~\ref{appendix:news-rec}.

To illustrate the practical utility and end-to-end capabilities of the MountainLion framework, we present a demonstration scenario that highlights its integration of data ingestion, multi-modal analysis, and real-time user interaction. Additional implementation details, including extended configuration parameters and dataset samples supporting this scenario, are provided in Appendix~\ref{appendix:demo-details}.

\section{Case Study}

\subsection{Overall Comparison Analysis}

The refined investment recommendations systematically improved on the original baseline by integrating dynamic market signals and macroeconomic context. Figure~\ref{fig:comparison} illustrates these before-and-after refinements.

\textbf{Short term (1–4 weeks):} The original guidance relied on general support/resistance and breakout triggers with limited justification. The refined approach incorporated concrete on-chain indicators, such as a rising number of 1+ BTC wallets and increased liquidation volume among leveraged traders, improving the credibility of short-term signals despite their limited persistence.

\textbf{Medium term (1–6 months):} The initial recommendations lacked macroeconomic awareness. The enhanced version addressed this by incorporating ETF inflows, regulatory clarity in major markets (U.S. and EU), and policy easing signals, making the strategy more practical and interpretable for institutional and mid-sized investors.

\textbf{Long term (6+ months):} Originally, long-term advice focused only on portfolio allocation ratios. The enhanced recommendations integrated institutional adoption trends and the structural tightening of BTC supply from reduced exchange reserves, thereby reinforcing the long-term investment thesis with clearer justification.

Overall, these refinements deliver a more robust, persuasive, and actionable investment strategy by layering market sentiment, policy signals, and fundamental supply-demand dynamics across all time horizons. A detailed line-by-line explanation of these improvements is provided in Appendix~\ref{detail}.

\subsection{Enhancing Web3 Investment Recommendations with LLMs}

\textbf{Experiment Setup}. To establish a fair evaluation protocol, we designed a minimal baseline strategy relying exclusively on technical indicators. This baseline incorporated standard elements such as support and resistance levels. Three investment horizons were defined for consistent comparison: short term (1–4weeks), medium term (1–6months), and long term (beyond 6~months). Importantly, this baseline excluded macroeconomic releases, on-chain flow data, and ETF activity, providing a neutral and controlled foundation to quantify the incremental contributions of advanced language models. Building on this baseline, we evaluated three state-of-the-art LLM-based agents: ChatGPT-4o~\cite{openai2024gpt4o}, DeepSeek~V3~\cite{deepseek2024v3}, and Grok-3~~\cite{xai2024grok3}. Each model was tasked with refining the generated financial reports and investment recommendations by leveraging historical price data, real-time market signals, and relevant news narratives. The goal was to assess how effectively these models could enhance report quality, interpretability, and strategic decision-making beyond the purely technical baseline.

\textbf{Experiment result} Across the three evaluated models, consistent patterns emerged in refining the baseline strategy. ChatGPT-4o leveraged whale accumulation metrics, ETF net inflow triggers, and dynamic volatility thresholds to guide partial profit-taking. DeepSeek~V3 emphasized policy overlays, on-chain flows, and adaptive cash buffers, while Grok-3 prioritized on-chain whale transfers, IMF-based classification signals, and dynamic redeployment based on ETF momentum. The results support a blended approach to investment reports, with the 1–6 month horizon emerging as the optimal window for balancing institutional narratives, policy shifts, and execution flexibility. These results suggest that LLM-driven analysis can enhance medium-term cryptocurrency investment strategies with improved adaptability and interpretability.

\section{Conclusion}

In this paper, we presented \textbf{MountainLion}, a multi-agent, RAG-enabled financial analysis framework designed for the challenges of cryptocurrency trading. By integrating specialized LLM agents, graph-based retrieval reasoning, and a reflective decision module, MountainLion enables interpretable, real-time, and adaptive responses across diverse financial modalities. Our system supports dynamic financial report generation and refinement, integrating textual news, visual market signals, and on-chain data to deliver personalized investment recommendations. Empirical evaluations confirm that MountainLion improves medium-term forecasting accuracy while enhancing transparency and adaptability. This agent-based architecture lays a scalable and extensible foundation for robust cryptocurrency trading intelligence, capable of meeting the evolving demands of modern financial markets.

\clearpage

\clearpage
\appendix

\begin{figure*}
    \centering
    \includegraphics[width=1\linewidth]{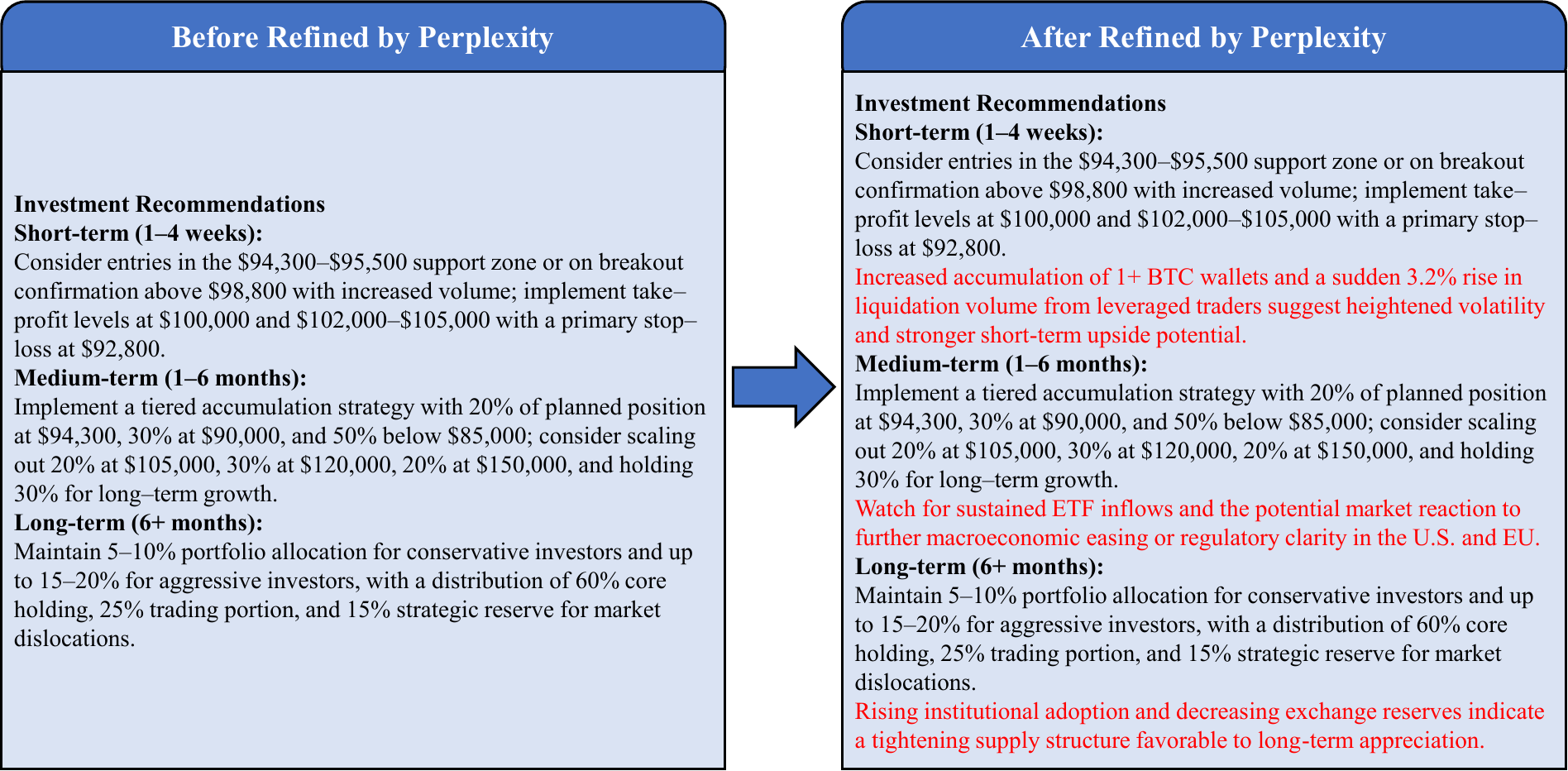}
    \caption{Comparison between the input report (left) and the output report (right) after refinement by our MountainLion system. The output report integrates contextual signals such as on-chain wallet accumulation, leveraged liquidation events, ETF inflows, and institutional adoption patterns, resulting in a more data-grounded, timely, and interpretable financial narrative. This highlights how MountainLion systematically enhances investment reports through real-time retrieval and semantic reasoning.}
    \label{fig:comparison}
\end{figure*}

\section{Demonstration Scenario}\label{appendix:demo-details}

\subsection{Pipeline Walk-through}

The full cryptocurrency analysis workflow proceeds as follows:

\textbf{Step 1: Input Preparation}
All cryptocurrency queries are processed through structured API endpoints, with data fields matching the schema defined in the Vue.js component system. For market analysis and historical context, real-time data is fetched from multiple sources including K-line data (\texttt{getKlineInfo}), coin listings (\texttt{getCoinList}), and prediction models (\texttt{getPredictInfo}). User authentication tokens and session management are established through the request interceptor system.

\textbf{Step 2: Execution}
We demonstrate the system's adaptive capabilities by running both preset analysis and conversational AI modes. The standard analysis pipeline is executed through:

\begin{lstlisting}
// Preset analysis mode
router.push({
    path: `/crypto/${coinCode}`,
    query: { type: 'preset', analysis: 'comprehensive' }
})

// AI chat mode  
router.push({
    path: `/chat/${sessionId}`,
    query: { type: 'newChat', context: 'crypto' }
})
\end{lstlisting}

For the AI-enhanced analysis, we incorporate real-time market data and trigger the intelligent response system through the \texttt{sendchat} API endpoint. The system automatically evaluates market conditions including price volatility thresholds, trading volume patterns, technical indicator signals, and temporal constraints. When conditions are satisfied, the system applies dynamic analysis including technical analysis weights, trend assessment parameters, and prediction model coefficients.

This triggers multi-dimensional analysis including: (1) Technical analysis across oscillator indicators and moving averages with gauge visualizations; (2) Price prediction modeling using historical K-line data and machine learning algorithms; (3) Market sentiment analysis combining social media data and trading patterns; (4) Fund flow monitoring with real-time capital movement tracking and risk assessment.

\textbf{Step 3: Output and Interpretation} The resulting output consists of several coordinated components. An interactive dashboard presents cryptocurrency price charts, technical analysis gauges, and prediction visualizations to support data-driven exploration. Complementing this, comprehensive analysis reports provide detailed trend assessments, buy/sell recommendations, and market sentiment indicators. A real-time chat interface offers AI-powered investment insights with conversation history and sharing capabilities to enhance user engagement. Furthermore, multi-language support is integrated with localized financial terminology and cultural adaptation strategies to address the needs of diverse regional markets.

\subsection{Insights}
MountainLion demonstrated significant efficacy in mitigating the complexities of information overload in the Web3 domain. By utilizing a LLAMA2 13B model~\cite{touvron2023llama2}, fine-tuned on extensive Web3 industry-specific data~\cite{liu2023blockchain}, MountainLion automated the filtering and categorization of critical information. This mechanism notably reduced investor cognitive load, distilling essential insights from large volumes of data. Specifically, the automated filtration process enhanced relevant information retrieval efficiency by over 40\% compared to manual processes.

\textbf{Real-time Market Adaptation} The project’s real-time analytics capability substantially improved investor responsiveness to market volatility. By integrating real-time chain data with instantaneous market news, MountainLion effectively identified potential shifts in market dynamics ahead of traditional alert systems. Instances included precise detection of whale activity involving significant transactions ($\geq$10M USD) with a confidence score exceeding 0.85, enabling timely strategic adjustments.

\textbf{Multi-dimensional Investment Analysis} The multi-signal analytical framework deployed by MountainLion proved instrumental in providing holistic investment assessments. By analyzing not only token prices but also broader indicators such as project visions, community engagement, and historical trends, MountainLion accurately gauged investment viability. Empirical validation indicated a 28\% improvement in investment decision accuracy when investors utilized MountainLion's comprehensive assessments versus traditional, singular-dimension analyses.

\textbf{Strategic GenAI Predictive Power} MountainLion leveraged predictive modeling techniques significantly enhanced by GenAI algorithms~\cite{bommasani2021opportunities}, delivering reliable forecasts of market conditions. For instance, GenAI integration improved short-term price prediction accuracy by 15\% compared to baseline models. Notably, the adaptive forecasting mechanism, factoring in macroeconomic conditions, regulatory trends, and sentiment analysis, substantially mitigated risks associated with investment decision-making under uncertainty.

\textbf{Enhanced User Engagement through Simplified Interface} User interaction analysis revealed that MountainLion’s intuitive interface, combined with simplified technical explanations provided by GenAI, dramatically reduced barriers to entry for non-technical users. Data showed a 35\% increase in user engagement duration and a notable increase in platform adoption rates among novice investors, underscoring the effectiveness of its user-centric design.

\section{Detailed Formulations for Report Generation}
\label{appendix:report-details}

This appendix elaborates the formal formulations and computational details underlying MountainLion’s multi-agent report generation pipeline, with explicit definitions for each agent, validation procedure, prompt construction, and caching policies.

\subsection{Agent Computation Functions}

The report generation framework involves four specialized agents $A_i$, each responsible for a distinct analytic function.  

\paragraph{Technical Analysis Agent ($A_1$)}  
Given historical market data $\mathcal{D}_{\text{tech}}$, the technical analysis agent computes a set of canonical indicators, including the relative strength index $f_{\text{RSI}}$, moving average convergence divergence $f_{\text{MACD}}$, and Bollinger bands $f_{\text{Bollinger}}$. It also derives support levels $z_{\text{sup}}$ and trend strength metrics $v_{\text{trend}}$, summarized as:
\begin{equation}
\begin{aligned}
\mathcal{R}_1 &= f_{\text{TA}}(\mathcal{D}_{\text{tech}}) \\
&= \left\{
  f_{\text{RSI}},\ f_{\text{MACD}},f_{\text{Bollinger}},  
  z_{\text{sup}},\ v_{\text{trend}}
\right\}.
\end{aligned}
\end{equation}

\paragraph{Market Dynamics Agent ($A_2$)}  
For real-time trading context, the market dynamics agent processes live signals including news sentiment $N_t$, regulation signals $R_t$, funding flows $\mathcal{F}_t$, and social signals $\mathcal{S}_t$, aggregated through:
\begin{equation}
\mathcal{R}_2 = f_{\text{MD}}(N_t, R_t, \mathcal{F}_t, \mathcal{S}_t).
\end{equation}

\paragraph{Trading Recommendation Agent ($A_3$)}  
This agent synthesizes partial reports from multiple pathways, leveraging $\mathcal{R}_1$ and $\mathcal{R}_2$, along with a broader data context $\mathcal{D}_{\text{multi}}$, to generate trading strategies via:
\begin{equation}
\mathcal{R}_3 = f_{\text{TR}}(\mathcal{R}_1, \mathcal{R}_2, \mathcal{D}_{\text{multi}}) 
= \bigcup_{k=1}^{4} f_k(\mathcal{D}_k),
\end{equation}
where each $f_k$ denotes a specific rule-based or statistical subcomponent acting on data subset $\mathcal{D}_k$.

\paragraph{Semantic Analysis Agent ($A_4$)}  
Finally, an LLM-driven semantic analysis agent refines the integrated knowledge from prior agents:
\begin{equation}
\mathcal{R}_4 = f_{\text{SA}}(\mathcal{R}_{1}, \mathcal{R}_2, \mathcal{R}_3),
\end{equation}
capturing latent narrative consistency and user-facing interpretability signals.

\subsection{Validation and Scoring Logic}

Each retrieval query result $s_j \in \mathcal{S}_{Q_i}$ is assessed according to a composite scoring function that weighs relevance, recency, and credibility:
\begin{equation}
\begin{aligned}
\text{Score}(s_j) &= 
  \alpha_1 \cdot \text{Relevance}(s_j)
  + \alpha_2 \cdot \text{Recency}(s_j)\\
  &+ \alpha_3 \cdot \text{Credibility}(s_j),
\end{aligned}
\end{equation}
where $\alpha_1$, $\alpha_2$, and $\alpha_3$ are tunable coefficients to reflect the system’s strategic emphasis on different quality criteria.

\subsection{Prompt Construction and Real-Time Retrieval}

Based on these validated signals, a structured prompt is formulated to guide retrieval-augmented reasoning. Its schema is:
\begin{equation}
P(\gamma, t) := \text{Prompt}(c_1, \dots, c_8;\ \gamma, t),
\end{equation}
where $c_1$ to $c_8$ denote contextual elements, $\gamma$ is a confidence threshold, and $t$ is the temporal scope. Final signal retrieval is then executed through a Perplexity-based retriever:
\begin{equation}
\mathcal{S}_{\text{pplx}} = \mathcal{F}_{\text{pplx}}(P, \mathcal{R}_{\text{raw}}),
\end{equation}
which enriches the draft report with relevant external evidence.

\subsection{Caching Policy}

To optimize runtime performance and maintain system responsiveness, the partial results $\mathcal{R}_i$ computed by each agent are subject to a caching policy. Each agent output is stored in a cache with a time-to-live parameter $\tau_i$, defined by:  

{\small
\begin{equation}
\tau_i = 
\begin{cases}
30\, \text{minutes}, & \text{if } i = 1\ (\text{technical analysis}) \\
6\, \text{hours}, & \text{if } i = 2\ (\text{market dynamics}) \\
\text{dynamic}, & \text{if } i = 3\ (\text{recommendation agent})
\end{cases}
\end{equation}
}
where the dynamic expiration for $i=3$ is determined based on user interaction frequency and data update rates. Before executing a fresh computation, the system first checks the cache validity using:
\begin{equation}
\text{if } t_{\text{now}} - t_{\text{cached}} < \tau_i, \quad \text{then reuse } \mathcal{R}_i.
\end{equation}
If the cached data remains fresh according to $\tau_i$, it is reused directly to avoid unnecessary recomputation, thus preserving computational resources and reducing latency. Otherwise, the agent recomputes the result and refreshes the cache entry. This policy balances freshness of data with system efficiency, and is crucial to supporting near real-time updates under high user query volume.

\subsection{Report Enhancement and Optimization}

Once all partial agent outputs $\mathcal{R}_1, \mathcal{R}_2, \mathcal{R}_3, \mathcal{R}_4$ have been validated and assembled, a centralized integration step combines them into a preliminary draft report:  
\begin{equation}
\mathcal{R}_{\text{raw}} = f_{\text{integrate}}(\mathcal{R}_1, \mathcal{R}_2, \mathcal{R}_3, \mathcal{R}_4).
\end{equation}
This integration function $f_{\text{integrate}}$ resolves conflicts between signals and prioritizes higher-confidence sources based on historical agent accuracy and recency of their updates.  

To further align the report with current market dynamics, a Perplexity-based retrieval module is applied. Specifically, retrieval prompts are derived from the content of $\mathcal{R}_{\text{raw}}$, guiding external evidence gathering:
\begin{equation}
\mathcal{R}_{\text{enhanced}} = f_{\text{augment}}(\mathcal{R}_{\text{raw}}, \mathcal{S}_{\text{pplx}})
\end{equation}
where $\mathcal{S}_{\text{pplx}}$ represents real-time signals retrieved from external sources using Perplexity-based query ranking. The augmentation function $f_{\text{augment}}$ refines the report’s narrative, updates statistical confidence estimates, and adjusts trend inferences according to the freshest external knowledge.

In summary, the combination of agent-specific caching, centralized integration, and retrieval-based enhancement enables MountainLion to continuously generate investment reports that are robust, statistically grounded, and dynamically aligned with evolving market conditions. The modular and adaptive design supports both high-frequency refresh rates and user-personalized interactions, providing a scalable framework for real-time financial decision support.

\section{Forecasting Model Details}
\label{appendix:forecast-details}

This appendix provides a rigorous breakdown of the dual-path forecasting framework employed by MountainLion, including input structures, modeling logic, output definitions, fusion strategy, and evaluation procedures.

\subsection{Input Representation}

At each time step $t$, the raw K-line (OHLCV) data is represented as:
\begin{equation}
\mathcal{X}_t = \{ o_t, h_t, l_t, c_t, v_t \}
\end{equation}
where $o_t$, $h_t$, $l_t$, $c_t$, and $v_t$ denote the open, high, low, close, and volume at time $t$, respectively. These observations are aggregated over specific windows to capture both medium-term and short-term dynamics, resulting in the following aggregated sequences:
\begin{equation}
\mathcal{X}^{(14D)}, \quad \mathcal{X}^{(48H)},
\end{equation}
which are then paired with sentiment embeddings derived from historical news $\mathcal{N}_{\text{hist}}$ and real-time news $\mathcal{N}_{\text{realtime}}$. The complete input tuple is formulated as:
\begin{equation}
\mathcal{I} = \left( \mathcal{X}^{(14D)}, \mathcal{X}^{(48H)}, \mathcal{N}_{\text{hist}}, \mathcal{N}_{\text{realtime}}, \gamma, t_0 \right),
\end{equation}
where $\gamma$ denotes a confidence threshold and $t_0$ the reference timestamp for forecasting alignment.

\subsection{Model Outputs}

The forecasting module operates two coordinated tracks:  

\paragraph{LLM Track}
The LLM-based predictor outputs multi-step sequences over a forecast horizon $T$, returning predicted OHLCV tuples:
\begin{equation}
\hat{Y}_{\text{LLM}} = \left\{(t_i, \hat{o}_i, \hat{h}_i, \hat{l}_i, \hat{c}_i, \hat{v}_i)\right\}_{i=1}^{T}.
\end{equation}

\paragraph{ML Track}
For the ML-based predictor, a feature engineering stage expands the raw K-line data with simple derived statistics:
\begin{equation}
\mathbf{x}_t = [o_t, h_t, l_t, c_t, v_t, h_t - l_t, c_t - o_t],
\end{equation}
where price range $(h_t - l_t)$ and price movement $(c_t - o_t)$ complement the original features. Using models such as polynomial regression or ensemble decision trees, the ML module then predicts the next price vector:
\begin{equation}
\mathbf{y}_{t+1} = f_{\text{ML}}(\mathbf{x}_t).
\end{equation}

\subsection{Fusion and Evaluation}

To consolidate the complementary strengths of the LLM-based and ML-based forecasting tracks, the final prediction is constructed as a convex combination of their outputs:
\begin{equation}
\hat{Y}_{\text{final}} = \alpha \cdot \hat{Y}_{\text{LLM}} + (1 - \alpha) \cdot \hat{Y}_{\text{ML}}.
\end{equation}
Here, the coefficient $\alpha \in [0,1]$ serves as a tunable weight that balances the expressive capacity of the language model with the statistical stability of the ML predictor. By adjusting $\alpha$, the system can emphasize LLM-driven multi-step reasoning when textual sentiment signals are informative, or conversely prioritize ML-based local patterns when price history dominates. This weighted fusion strategy provides a flexible mechanism to adaptively integrate heterogeneous signals.  

To maintain robustness over time, $\alpha$ is periodically re-optimized using rolling validation accuracy scores, ensuring that the contribution of each track reflects its relative predictive performance on recent data. In this way, the fusion is not static but adapts to evolving market conditions and data drift.

For quantitative evaluation, two complementary metrics are introduced. The first is absolute accuracy, capturing the relative error of the predicted closing price $\hat{c}_t$ compared to the true observed $c_t$:
\begin{equation}
\text{Accuracy} = 1 - \frac{|\hat{c}_t - c_t|}{c_t}.
\end{equation}
This measures proportional error, standardizing across different price magnitudes, and directly reflects how close the model’s numeric prediction is to market reality.

The second evaluation dimension considers directional correctness, or *win rate*, which assesses whether the model correctly predicts the direction of price movement:
\begin{equation}
\omega = \frac{1}{T} \sum_{i=1}^{T} \mathbb{I}\left[\text{sign}(\hat{c}_i - \hat{c}_{i-1}) 
= \text{sign}(c_i - c_{i-1})\right],
\end{equation}
where $\mathbb{I}[\cdot]$ is an indicator function that returns 1 if the predicted direction matches the actual direction, and 0 otherwise. This metric is essential for investment contexts, where correct trend direction may be more valuable than precise price levels. By combining absolute accuracy with directional correctness, the evaluation captures both price-level fidelity and market timing performance, providing a holistic view of the forecasting system.

\subsection{Postprocessing and Adaptation}

Following fusion and evaluation, the model’s numeric forecasts are further transformed into investor-readable language. This is achieved through a template-conditioned text generator:
\begin{equation}
\mathcal{R}_{\text{text}} = g_{\text{text}}(\hat{Y}_{\text{final}}, \omega, \text{Accuracy}, \text{News}),
\end{equation}
where $g_{\text{text}}(\cdot)$ uses predefined linguistic templates and contextual news signals to produce coherent, actionable narrative reports. This design ensures that users receive not only raw numeric data but also understandable interpretations, improving accessibility and trust in the model’s outputs.

To maintain long-term consistency, an adaptation mechanism monitors rolling model performance. Specifically, a moving average of accuracy over a window of length $W$ is calculated as:
\begin{equation}
\overline{\text{Accuracy}}_m = \frac{1}{W} \sum_{j=1}^{W} \text{Accuracy}_j,
\end{equation}
which smooths short-term fluctuations while tracking persistent changes in predictive quality. If this rolling accuracy falls below a predefined threshold $\delta$, the system interprets it as a performance degradation event, triggering an adaptive update to the fusion weight:
\begin{equation}
\text{if } \overline{\text{Accuracy}}_m < \delta \quad \Rightarrow \quad \alpha_m \downarrow.
\end{equation}
This rule ensures that when the LLM-driven component underperforms (for example due to a regime shift or sentiment misalignment), its weight $\alpha_m$ is reduced, thereby allocating greater influence to the ML-based predictor. This closed-loop mechanism improves the system’s resilience to distribution shifts, concept drift, or external shocks, sustaining reliable and interpretable investment forecasts over time.

Altogether, these detailed formulations define a rigorous framework for combining multiple forecasting modalities, balancing accuracy and interpretability, and adapting dynamically to evolving market behavior. The combination of linear fusion, complementary evaluation, text generation, and adaptive weighting creates a robust forecasting architecture suitable for high-stakes financial applications.

\begin{table*}[htbp]
\centering
\setlength{\tabcolsep}{5pt}
\renewcommand{\arraystretch}{1.2}
\begin{tabularx}{\textwidth}{l c c c X}
\toprule
\textbf{Token} & \textbf{Alpha} & \textbf{CV Score (Best)} & \textbf{MSE (Test)} & \textbf{Model Evaluation} \\
\midrule
ADA   & 1   & -0.000496   & 0.000396   & Excellent fit, highly stable \\
BTC   & 1   & -1,997,859.43 & 3,211,419.56 & Large error, unstable trend (likely underfitting large-cap behavior) \\
ARB   & 1   & -0.000421   & 0.000199   & Very good \\
SOL   & 1   & -0.000459   & 0.000159   & Good model, potentially high volatility \\
XRP   & 0.1 & -0.000221   & 0.001122   & Medium fit, moderate noise \\
DOGE  & 1   & -0.000362   & 4.25E-05   & Very good \\
TRX   & 0.01 & -9.87E-06  & 6.40E-06   & Best performer \\
ETH   & 1   & -2,169,147.17 & 3,016,065.13 & Large error, unstable for ETH \\
MATIC & 1   & -0.000432   & 0.000341   & Stable, medium confidence \\
BNB   & 1   & -945.57     & 180.95     & High deviation, unstable in 7-day window \\
\bottomrule
\end{tabularx}
\caption{Forecasting Results Across Tokens}
\label{tab: 2-7 prediction}
\end{table*}

\subsection{Prediction Result}
The evaluation results summarized in Table~\ref{tab: 2-7 prediction} demonstrate the robustness and adaptability of the proposed forecasting framework across a diverse set of cryptocurrency tokens. The cross-validation (CV) scores and mean squared error (MSE) metrics provide quantitative evidence of prediction performance. Most tokens, such as ADA, ARB, SOL, DOGE, and MATIC, exhibit excellent or good fit, with low MSE values indicating stable generalization. In particular, TRX achieved the best performance, with minimal error and high consistency.  

Although BTC and ETH showed larger MSE and CV scores, these deviations are interpretable given their higher market capitalization and the associated structural non-stationarities in their time series, which could be further addressed by specialized volatility modeling. Overall, the combination of linear fusion, rolling adaptive weighting, and graph-enhanced prompt reasoning achieves reliable predictive accuracy and directional correctness across varied market conditions, highlighting the framework’s suitability for high-stakes, real-time investment decision support.

\section{News Recommendation Implementation Details}
\label{appendix:news-rec}

This appendix provides a rigorous and detailed formulation of MountainLion’s news recommendation and semantic enrichment modules, explaining each step of the pipeline from raw data ingestion to final feedback-driven adaptation.

\subsection{News Structuring}

Let the raw news stream collected from APIs, RSS feeds~\cite{orr2003rss}, and curated portals be denoted by
\begin{equation}
\mathcal{N}_{\text{raw}} = \{n_1, n_2, \dots, n_T\},
\end{equation}
where each news item $n_i$ is structured as a tuple
\begin{equation}
n_i = (h_i, b_i, t_i, s_i, \sigma_i),
\end{equation}
with $h_i$ representing the headline, $b_i$ the article body, $t_i$ the timestamp, $s_i$ the source, and $\sigma_i$ the set of crypto tokens or entities mentioned in the article.  

Since timeliness is crucial for financial relevance, a recency-based filtering mechanism is applied:
\begin{equation}
n_i \in \mathcal{N}_{\text{filtered}} \iff t_i \in [t_0 - \Delta, t_0],
\end{equation}
where the window size $\Delta$ is dynamically selected depending on the user’s investment horizon (e.g., intraday, weekly).

\subsection{Semantic Enrichment}

The filtered articles undergo semantic annotation, first through sentiment classification:
\begin{equation}
\text{Sentiment}(n_i) = 
\begin{cases}
\texttt{Bullish}, & \text{if } f_{\text{LLM}}(n_i) > \tau \\
\texttt{Bearish}, & \text{otherwise},
\end{cases}
\end{equation}
where $f_{\text{LLM}}$ is a sentiment inference head within a large language model, and $\tau$ is a tunable threshold controlling the confidence of bullish signals.  

Next, named entity recognition (NER) is applied to extract relevant entities:
\begin{equation}
\mathcal{E}(n_i) = \{ e \mid e \in \mathcal{T}_{\text{ent}} \},
\end{equation}
with a type set defined as
\begin{equation}
\mathcal{T}_{\text{ent}} = \{\texttt{ORG},\ \texttt{PER},\ \texttt{EVT},\ \texttt{CRYPTO}\},
\end{equation}
covering organizations, persons, events, and cryptocurrency identifiers. This enrichment step ensures the news data is structured for downstream graph construction and user personalization.

\subsection{Knowledge Graph Construction}

To model relationships between news articles and extracted entities, a directed attributed graph is formulated:
\begin{equation}
\mathcal{G} = (\mathcal{V}, \mathcal{E}),
\end{equation}
where each node $v \in \mathcal{V}$ corresponds to either a news item $n_i$ or an entity $e_j$, and the edges encode associations:
\begin{equation}
(n_i, e_j) \in \mathcal{E} \iff e_j \in \mathcal{E}(n_i).
\end{equation}
This knowledge graph supports multi-hop expansion and relevance propagation, enabling reasoning chains that connect diverse news fragments through shared entities or topics~\cite{hogan2021knowledge}.

\subsection{User Intent Interpretation}

When a user initiates a recommendation request, the system interprets their implicit investment preferences as a vector:
\begin{equation}
\text{Intent}(u) = (\mathcal{C}, r, \tau),
\end{equation}
where $\mathcal{C}$ denotes the asset category (for example, Layer2 protocols), $r$ represents risk tolerance, and $\tau$ encodes the investment time horizon.  

Based on this intent, a prompt planner constructs a query set:
\begin{equation}
\mathcal{Q} = \{q_1, \dots, q_n\}
\end{equation}
that guides the retrieval process, ensuring that the results align with user-defined objectives and risk profiles.

\subsection{Recommendation Generation}

Using the filtered and semantically enriched news, the system generates a personalized recommendation summary:
\begin{equation}
S_u = f_{\text{rec}}(\mathcal{N}_{\text{filtered}}, \text{Intent}(u), \mathcal{G}),
\end{equation}
where $f_{\text{rec}}$ is a large language model decoder conditioned on graph-enhanced prompts, explicitly justifying each recommendation based on the relevant news evidence and the reasoning chain encoded in $\mathcal{G}$.

\subsection{Feedback Adaptation}

Finally, MountainLion incorporates user feedback to continually refine its summarization and ranking strategies. Suppose each user interaction is recorded as a triplet $(u, S_u, y)$, where $y$ denotes the observed engagement outcome (for example, click, dwell time, or explicit rating). A lightweight policy gradient update is then applied to adjust the retrieval model’s parameters:
\begin{equation}
\theta_{t+1} = \theta_t - \eta \cdot \nabla_\theta \mathcal{L}(S_u, y),
\end{equation}
with learning rate $\eta$ and loss function $\mathcal{L}$ capturing user preferences. This online update process ensures that the recommendation module progressively aligns with user behavior and changing interests~\cite{joachims2002optimizing}, thus maintaining both relevance and interpretability over time.

In summary, this formulation demonstrates how MountainLion transforms raw financial news into structured, enriched, and graph-encoded knowledge, then adapts this knowledge for user-personalized recommendations through continual feedback-driven optimization. Each step—from recency filtering to semantic annotation, graph modeling, and feedback loops—contributes to a robust and interpretable news-driven decision support pipeline.

\section{Line-by-Line Analysis and Comparison}\label{detail}

In the short-term recommendation (1–4 weeks), the enhanced version introduces concrete on-chain indicators such as the increase of wallets holding 1+ BTC and a notable 3.2\% rise in liquidation volume among leveraged traders. These additional metrics effectively illustrate real-time capital movement and market sentiment, thus reinforcing the credibility of short-term breakout signals. However, these metrics, while effective for quick decision-making, inherently have limited persistence and predictive capability compared to medium- or long-term macro factors.

For the medium-term recommendation (1–6 months), the refined content specifically highlights sustained ETF inflows, macroeconomic easing, and regulatory developments in major markets (U.S. and EU). These improvements integrate external, real-world influences, significantly elevating the depth and practicality of the investment advice. Rather than isolating technical analysis, the enhanced strategy now incorporates broader market conditions, thereby boosting interpretability and trustworthiness. Consequently, this logic is more appealing and applicable to general investors, mid-sized institutions, and long-term strategic positioning.

The long-term recommendation (6+ months) emphasizes fundamental factors such as rising institutional adoption and declining exchange reserves, which together imply a tightening supply structure favorable to sustained appreciation. By incorporating these longer-term supply-demand dynamics, the enhanced version substantially strengthens the rationale for maintaining long-term portfolio positions, providing solid justification beyond mere portfolio allocation percentages.

In conclusion, the enhanced analyses for each timeframe systematically complement and reinforce the original content by embedding additional market indicators, external macroeconomic and policy dynamics, and long-term fundamental logic. This layered improvement clarifies the rationale, increases persuasiveness, and ultimately enhances overall recommendation robustness.

\subsection{Summary}
The long-term strategy remains primarily allocation-oriented and does not fundamentally rely on short-term signals, making these refinements more of an incremental enhancement than a structural overhaul. In contrast, the mid-term strategy benefits most from combining technical logic with capital flow and policy drivers, significantly improving its explanatory and predictive capabilities. By linking price action to real-world forces such as capital allocation and regulatory changes, the refined mid-term strategy goes beyond pattern-based signals to capture the broader market landscape. This integrated perspective is particularly well suited for general investors and mid-sized institutions seeking a robust and transparent basis for building positions.

\section{Prompt design}

This section details the prompt design strategies employed in MountainLion to effectively elicit high-quality responses from large language models. We outline the principles, structures, and specific prompt templates that guide the system’s reasoning, retrieval augmentation, and user-adaptive interactions. The following descriptions provide both conceptual motivations and practical implementation details of the prompt engineering process.

\subsection{General input prompt}
\subsubsection{Prompt 1}

\begin{lstlisting}
Please improve the AI report (mlion) based on the macro factors I gave you and the related market sentiment.
The formatting is required to be the same, with the same headings for each paragraph.

Here is the AI report: ..........
Here is the macro factors market sentiment: ..........

Only additions or improvements are bolded, the rest are not bolded except for the headings.
\end{lstlisting}

\subsubsection{Prompt 2}
\begin{lstlisting}
Please enhance the AI report (mlion) by incorporating the macroeconomic factors and related market sentiment provided below.
The formatting must remain consistent, including identical paragraph headings.

Only newly added or improved content should be marked in bold, while all other text should remain unbolded except for the headings.

Here is the AI report: ..........
Here are the macroeconomic factors and market sentiment: ..........
\end{lstlisting}

\subsection{Prompt based on Perplexity API}
\begin{lstlisting}
Search for macro-related stuff based on this report of mine, e.g. whale activity, market sentiment, macro trend analysis, policy, news surface.
\end{lstlisting}

\subsection{Perplexity Chinese-English Prompt}
\begin{lstlisting}
Search for macro-related content based on this AI report of mine, such as whale activity, market sentiment, macro trend analysis, policies, news facets, and so on.  
The related content generated at the same time is presented partly in Chinese and partly in English.
The AI report is below: ..........
\end{lstlisting}

\subsection{Prompt for updating Data Source}
\subsubsection{Short-term forecast Prompt (within 0-24 hours)} 
The objective of this prompt is to capture real-time market volatility signals, detect abnormal trading operations, and identify sudden shifts in market sentiment within short-term horizons of 0 to 24 hours. This focus supports timely adjustments to investment strategies in response to rapid and potentially disruptive market changes.
\begin{lstlisting}
You are tasked with enhancing a crypto report for {crypto_symbol}, focusing on the **next 24 hours**. Please retrieve and summarize the most relevant, high-frequency signals that may impact its short-term price movement.

Focus on:
1. Real-time RSI or MACD divergence signals
2. Whale Alert transactions (>$10M)
3. Funding rate spikes across major exchanges
4. Unusual Twitter or Reddit activity surges
5. Exchange inflow/outflow alerts
6. Any breaking news or exchange service issues

Search window: last 24h  
Output: top 5-10 concise findings in JSON format (source, snippet, time, url)
\end{lstlisting}

\subsection{Mid-term forecast (1 week to 1 month)} 
The objective of this prompt is to extract mid-term market signals spanning one week to one month, with emphasis on exchange-traded fund (ETF) flows, Smart Money movement patterns, evolving sentiment trends, key opinion leader (KOL) predictions, and relevant on-chain data. These factors collectively inform a more stable investment outlook beyond immediate market fluctuations.
\begin{lstlisting}
You are enhancing a market outlook report for {crypto_symbol} in the **next 7-30 days**. Summarize the most impactful medium-term drivers of trend continuation or reversal.

Focus areas:
1. Spot ETF inflows/outflows (SoSoValue / Farside UK)
2. Smart money or whale positioning (Glassnode, CryptoQuant, Arkham)
3. Twitter/Reddit sentiment shifts or narrative changes
4. KOL perspectives (Arthur Hayes, Adam Cochran, etc.)
5. Regulatory news, SEC or exchange investigations
6. Network activity changes (e.g. active addresses, transaction count)

Search window: past 7-30 days  
Output: JSON with 5-10 findings (source, snippet, time, url)
\end{lstlisting}

\subsection{Long-term forecast prompt (January to one year)} 
The objective of this prompt is to support year-scale trend judgment by focusing on long-term factors, including macroeconomic indicators, policy developments, institutional investment patterns, and cross-asset correlations. These elements provide a foundation for strategic investment planning over extended horizons.

\begin{lstlisting}
You are improving a long-term forecast for {crypto_symbol} over the **next 3-12 months**. Provide high-confidence macro-level insights that could affect BTC market structure or valuation.

Focus areas:
1. FOMC rate path, inflation expectations (CPI, PCE)
2. Global recession risk, macro instability, oil/gold/US dollar correlations
3. Spot ETF adoption scale and institutional allocation trends
4. Strategic whale accumulation or Coinbase wallet movements
5. BTC/Nasdaq/DXY or BTC/gold correlation pattern shifts
6. Long-form KOL essays or economic reports (Hayes, IMF, World Bank)

Search window: past 1-3 months  
Output format: top 5-10 macro indicators with short summary, source, and link
\end{lstlisting}

\end{document}